\documentclass[twocolumn,amsmath,amssymb,aps,prl]{revtex4}

\usepackage{graphicx}
\usepackage{dcolumn}
\usepackage{bm}
\begin{document}
\pagestyle{empty}

\title{Morphogen gradient from a noisy source}
\author{Jeremy L. England$^{\dagger}$}
\author{John Cardy$^{\dagger\ddagger}$}
\affiliation{ Oxford University, Rudolf Peierls Centre for Theoretical Physics, 1 Keble Road, Oxford, OX1 3NP, United
Kingdom$^{\dagger}$ }
\affiliation{All Souls College, Oxford, United Kingdom$^{\ddagger}$}
\date{\today}

\begin{abstract} We investigate the effect of time-dependent noise on the shape of a morphogen gradient in a developing
embryo.  
Perturbation theory is used to calculate the deviations from deterministic behavior 
in a simple reaction-diffusion model of robust gradient formation, and the results are confirmed by numerical
simulation. It is shown that such deviations can disrupt robustness for sufficiently high noise levels, and the
implications of these findings for more complex models of gradient-shaping pathways are discussed.
\end{abstract}

\maketitle

\begin{figure}
\resizebox{\columnwidth}{!}{
\includegraphics{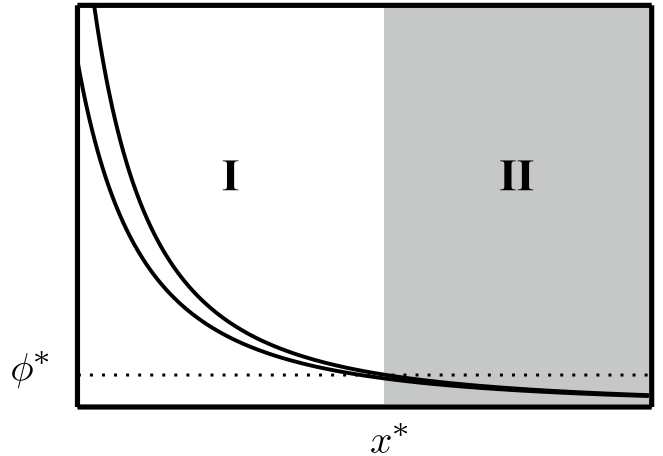} }
\caption{\label{fig:schem} Morphogen concentration gradients $\phi_{0}(x)$ are plotted from (\ref{eq:ss}) for two
different values of the production rate $\eta_{0}$. Where $\phi_{0}$ is greater than the threshold concentration
$\phi^{*}$, a cell fate of type I is induced (unshaded).  In contrast, $\phi_{0}<\phi^{*}$ induces a cell fate of type
II (shaded).  The process by which the gradient is shaped is said to be robust to $\eta_{0}$ because both curves
demarcate essentially the same location $x^{*}$ for the boundary between cell types.}
\end{figure}

Successful embryonic development requires that different regions within the embryo receive distinct sets of
developmental instructions. These instructions are often relayed by a morphogen, a signaling molecule that induces
different cell fates at different concentrations.  By establishing concentration gradients of various morphogens during
early stages of development, the embryo provides each cell with the positional information it needs for the proper
implementation of a body plan (Fig. \ref{fig:schem}) \cite{Teleman01}. 

A morphogen gradient must reliably demarcate precise and accurate boundaries between groups of cells despite substantial
genetic and environmental variability. Past studies have argued that this task might be accomplished in at least two
different ways.  First, it has been proposed that the downstream pathway that responds to a morphogen may have built-in
mechanisms for interpreting sharply-defined positional information from a gradient that varies quite substantially from
embryo to embryo \cite{Leibler02}.  Second, the shape of a gradient might itself be ``robust" to certain kinds of
interference. 
Eldar et al. have presented evidence that the shape of the BMP morphogen gradient may not be altered significantly by
several types of mutations in the pathway responsible for establishing the gradient \cite{Eldar02}.  Furthermore, their
theoretical investigations of dynamical models for several gradient-shaping pathways have uncovered steady-state
solutions that are effectively independent of certain kinetic parameters and initial or boundary conditions
\cite{Eldar02,Eldar03}.  Thus, there is some indication that 
a pathway's architecture could ensure that the shape of the resulting gradient did not depend on the rates of synthesis
or the concentrations of some or all components of the pathway. 

The theoretical studies of robustness mentioned above rely on deterministic reaction-diffusion equations that treat the
influxes of the components of a gradient-shaping pathway as constant, time-independent quantities. 
However, a growing number of experimental studies have demonstrated that gene expression at the cellular level is an
inherently ``noisy" process, during which the rates of synthesis of various gene products fluctuate substantially about
their average values over time \cite{Elowitz02,Ozbudak02,Blake03,Oshea04}. In the presence of such noise, a gradient
would 
be constantly driven away from its steady-state shape, and might even adopt a different shape, on average.   
Thus, the question of whether deterministic models can adequately capture the dynamics of morphogens that are, in
reality, produced and secreted stochastically
 is an open one whose answer may bear profoundly on our understanding of how a viable final product is assured during
the process of embryonic development.

This Letter introduces the novel dimension of time-dependent noise to the study of morphogen gradients. A perturbative
treatment of a simple reaction-diffusion model for gradient formation is used to calculate both the noise-averaged shift
of the system away from its ``robust" deterministic steady-state, and the system's fluctuations about this average. 
These analytical results are subsequently confirmed by a numerical simulation. 
Comparing our findings to the steady-state solution of the reaction-diffusion model, we find that deterministic
robustness can indeed be compromised by noise, but only when the intensity of noise is sufficiently high. Finally, we
argue that models of robust gradient formation that incorporate a more realistic number of components 
 should be even more susceptible to the effects of noise than the simple model investigated here.

 We begin by considering a single morphogen of local concentration $\phi(x,t)$ that is produced at the origin at a
constant rate $\eta_{0}$ and diffuses out to infinity along the positive $x$-axis with Fick's constant $D$.  
The morphogen accelerates its own rate of degradation such that it is removed from the system at a local rate of
$f\phi(x,t)^{2}$, which gives the equation of motion
\begin{equation}
\label{eq:grad}
\partial_{t}\phi=D~
\partial_{x}^{2}\phi-f \phi^{2}+\eta_{0}~\delta(x)
\end{equation}
Experimental evidence suggests that this mechanism of self-enhanced ligand degradation is a qualitatively correct
description of how the Wg gradient is established in fruit flies \cite{Eldar03}. 

When the gradient reaches a steady-state shape $\phi_{0}$, it will divide space into two regions: one where the
concentration of morphogen is higher than some threshold value $\phi^{*}$ and a cell fate of type I is induced, and one
where the concentration is lower and a cell fate of type II is induced (Fig. \ref{fig:schem}).  The boundary between the
two regions lies at the coordinate $x^{*}$ that satisfies $\phi_{0}(x^{*})=\phi^{*}$. 

The steady-state solution that satisfies $
\partial_{t}\phi_{0}=0$ is given by
\begin{equation}
\label{eq:ss}
\phi_{0}(x)=\frac{6
D}{f}\frac{1}{(x+\epsilon)^{2}}=\frac{6D}{f}\left[\frac{1}{x^{2}}-\frac{2}{x^{3}}\epsilon+\mathcal{O}(\epsilon^{2%
})\right]
\end{equation}
where $\epsilon = (12D^{2}/f\eta_{0})^{1/3}$.  The boundary coordinate is therefore given by
$x^{*}=\sqrt{6D/f\phi^{*}}-\epsilon\equiv x^{*}(0)-\epsilon$.  If $\epsilon\ll x^{*}(\epsilon)$, then $d\log
x^{*}/d\log\epsilon$ will be very small, meaning that relatively large percentage changes in $\epsilon$ will only
introduce minor percentage shifts in $x^{*}$.  Thus, so long as $\epsilon\ll x^{*}(0)$,
 the location of a cell type boundary specified as in Figure \ref{fig:schem} 
will be robust to the concerted shifts in $\eta_{0}$ that might result from genetic or environmental variation within a
population of embryos \cite{Eldar03}.  

\begin{figure}
\resizebox{\columnwidth}{!}{
\includegraphics{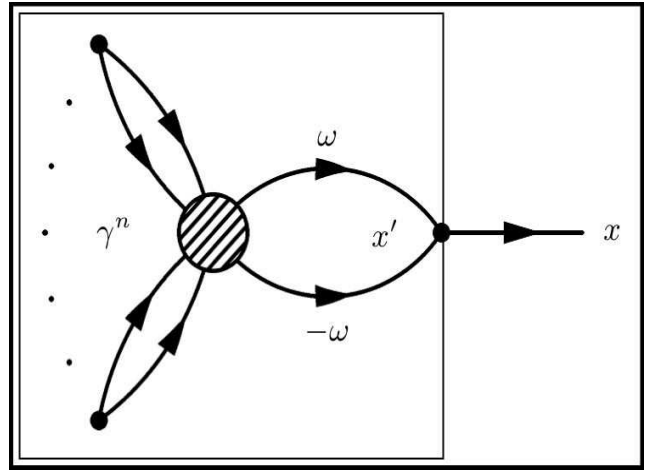} }
\caption{\label{fig:diag} The Feynman diagram for a general, $n$th-order process, with time increasing from left to
right. $n$ distinct fluctuation events at the origin propagate through space and interact with each other to produce a
positive contribution to $\Delta$ at $x'$ (inner box). 
The self-interaction of this variance summed over all points $x'$ then induces a non-zero $\langle\delta\phi\rangle$ at
$x$ (outer box).} 
\end{figure}

The preceding analysis operated under the assumption that the rate at which the morphogen is synthesized and secreted is
constant over time in a single embryo.  We now relax this assumption in order to study the consequences of introducing
time-dependent noise into the system.  The first step in doing this is to define new quantities
$\phi(x,t)=\phi_{0}(x)+\delta\phi(x,t)$ and $\eta(t)=\eta_{0}+\delta\eta(t)$.  A rescaling of space, time, noise, and
concentration in the interest of consolidating free parameters results in the new equation of motion
 \begin{equation}
 \label{eq:eqnmo}
\partial_{t}\delta\phi=
\partial_{x}^{2}\delta\phi-\frac{12 }{(x+\epsilon)^{2}}\delta\phi-\delta\phi^{2}+\delta\eta~\delta(x)
\end{equation}

We have computed the mean shift $\langle\delta\phi(x)\rangle$ and variance
$\Delta\equiv\langle\delta\phi(x)^{2}\rangle-\langle\delta\phi(x)\rangle^{2}$ (Here, $\langle\ldots\rangle$ denotes
averaging over all realizations of the noise $\delta\eta(t)$).  Since dimensional analysis requires that these two quantities 
depend on $\epsilon(\eta_{0})$, they are a means to gauge the impact that noise has on the robustness of the gradient to changes in
$\eta_{0}$. The noise itself is taken to satisfy $\langle\delta\eta(t)\rangle=0$ and $\langle\delta\eta(t)\delta\eta(t')\rangle=2\gamma\delta(t-t')$. 
A power-counting argument shows that the effects of more realistic noise correlations should not alter the behavior for
large $x$.

The quantities of interest can be calculated using perturbation theory.  The first step is to construct a Green's
propagator, which is composed of the eigenfunctions $\hat{\psi}_{\lambda}$ of the linear operator
$-\partial_{x}^{2}+12/(x+\epsilon)^{2}$ in the right-hand side of (\ref{eq:eqnmo}):
\begin{equation}
G(x_{1},x_{2};\omega)=\int_{0}^{\infty}
d\lambda\frac{\hat{\psi}_{\lambda}(x_{1})\hat{\psi}_{\lambda}(x_{2})}{-i\omega+\lambda}
\end{equation}
 The eigenfunctions may be approximated as
$\hat{\psi}_{\lambda}(x)=\sqrt{(x+\epsilon)/2}J_{7/2}[\sqrt{\lambda}(x+\epsilon)]$.  While terms proportional to Bessel
functions of the second kind are needed for exactness when $\epsilon\neq 0$, their contribution should be small for the
soft modes of the system that contribute most strongly to averages over the noise, particularly in the case of interest
where $\epsilon$ is small compared with the size of the system.

The iterative solution to (\ref{eq:eqnmo}) obtained by repeated substitution for the non-linear term can be expressed as
a sum over tree Feynman diagrams rooted at $x$, each edge corresponding to $G$, each node to the non-linear interaction,
and the end of each branch corresponding to an insertion of the noise. Averaging over the noise then has the effect of
pairing these ends in all possible ways. The $n$th-order contribution to $\Delta$ is represented in the inner box of
Figure
\ref{fig:diag}.  The two lines that converge on $x'$ must originate at vertices located at points $x_{1}$ and $x_{2}$,
 which can be assumed to be less than $x'$ for reasons that will be explained below.  Integration over $\omega$ and the
propagator eigenvalue for $G(x_{1},x')$ reveals that the $n$th-order diagram contains the integral
\begin{widetext}
\begin{equation}
\begin{aligned}
\label{eq:diag}
\ldots\frac{1}{4}\int d\lambda~
(x'+\epsilon)\sqrt{(x_{1}+\epsilon)(x_{2}+\epsilon)}J_{7/2}(\sqrt{\lambda}[x_{1}+\epsilon])J_{7/2}(\sqrt{\lambda}[x'%
+\epsilon])I_{7/2}(\sqrt{\lambda}[x_{2}+\epsilon])K_{7/2}(\sqrt{\lambda}[x'+\epsilon])
\end{aligned}
\end{equation}
\end{widetext} Assuming that $x'+\epsilon\gg x_{i}+\epsilon$ for all points $x_{i}$ that temporally precede $x'$ in Fig.
\ref{fig:diag}, the heavy damping from $K_{7/2}$ permits the substitution
$J_{7/2}(\sqrt{\lambda}[x_{1}+\epsilon])I_{7/2}(\sqrt{\lambda}[x_{2}+\epsilon])\sim \lambda^{7/2}$. The subsequent
integral over $\lambda$ in (\ref{eq:diag}) comes out the same regardless of the order in perturbation theory.  Thus, we
obtain $\Delta\sim 1/(x')^{8}$.

The above result was based on the assumption that $x'$ is always much larger than the spatial coordinates of vertices
located at earlier time coordinates.  Obviously, when one integrates over all possible vertex positions, one will be
forced to consider cases where $x_{1}\gg x'$.  However, free fluctuations constantly relax towards
$\delta\phi=0$ as they propagate through space, and their effects are clearly the strongest if they interact non-linearly
prior to undergoing most of this relaxation.
 Thus, by assuming that all interactions contributing to a diagram occur before propagation over any significant
distance takes place, we ignore processes that would have contributed only weakly to the final answer.

With $\Delta$ in hand, we now proceed to the outer box of Fig.
\ref{fig:diag} and calculate the average shift in $\delta\phi$ using
\begin{equation}
\label{eq:tad}
\langle\delta\phi(x)\rangle=\int dx'\Delta(x') G(x,x';\omega=0)
\end{equation}
Since
\begin{equation}
\label{eq:prop} G(x>x';0)=\frac{\sqrt{x~x'}}{7}\left(\frac{x'}{x}\right)^{7/2}
\end{equation}
the leading term in (\ref{eq:tad}) becomes $\langle\delta\phi(x)\rangle\sim 1/x^{3}$. It should be noted, however, that
the cubic power law is not a result of the specific form of the fluctuations $\Delta$ as long as they fall off
sufficiently rapidly with $x'$.  Rather, the observed decay would be brought in as part of any computation of
(\ref{eq:tad}) because of the cubic spatial dependence of $G(x>x';0)$. Thus, it seems likely that little was sacrificed
in generality by assuming a white spectrum for the noise.
\begin{figure}[t]
\resizebox{\columnwidth}{!}{
\includegraphics{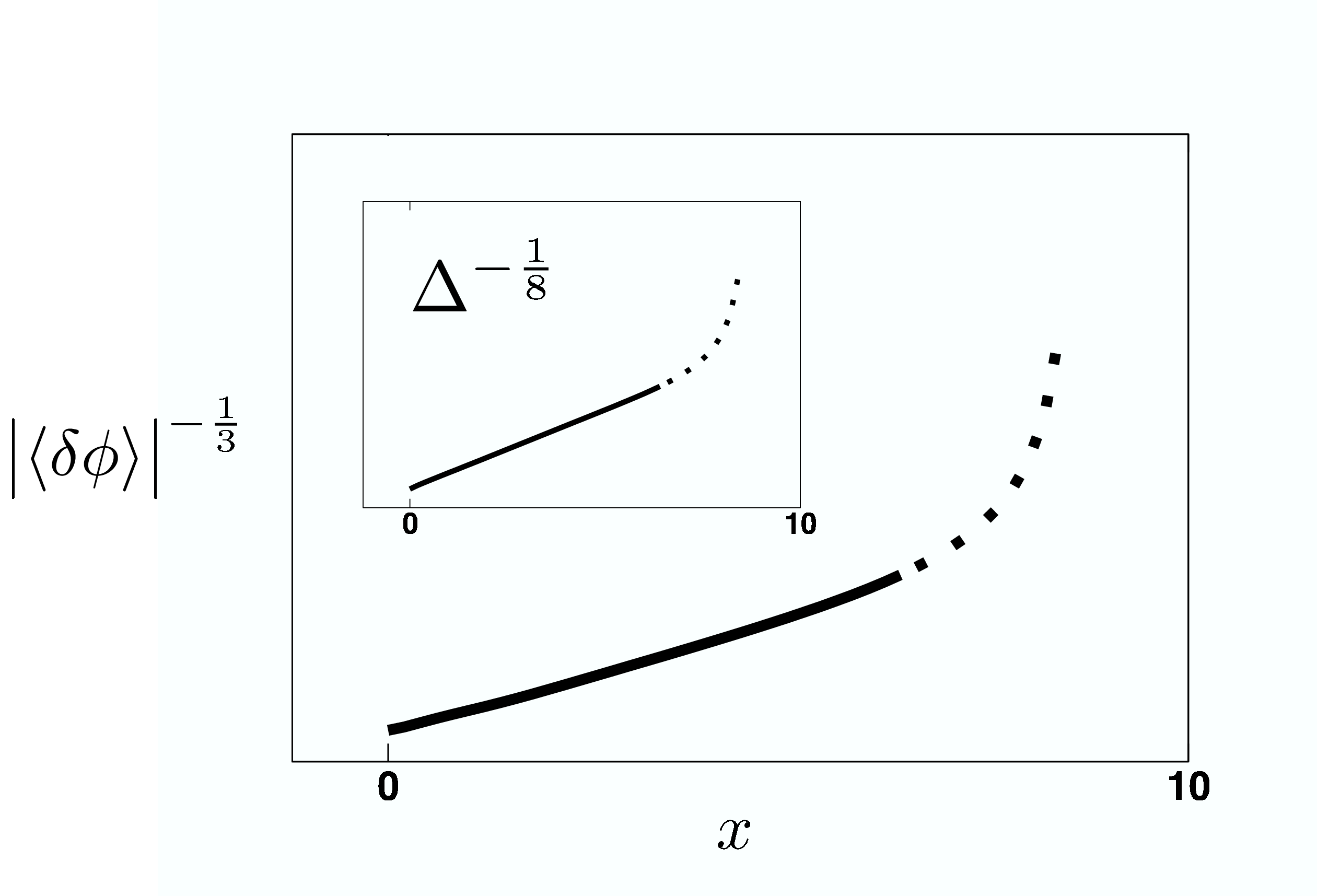} }
\caption{\label{fig:sim} The equation of motion (\ref{eq:eqnmo}) was solved numerically for $\Delta t = .001$, $\Delta
x=.2$, $f=5$, $\gamma=.2$, and $D=1$ on $0<x<8.8$.  Boundary conditions were imposed such that
$\partial_{x}\delta\phi\big|_{x=0}=\delta\phi(8.8)=0$.  Here, we plot the inverse cube root of the magnitude of the
average shift in $\delta\phi$, along with the inverse eighth root of the fluctuation of the shift about its average
value (inset). 
Regression analysis of the solid lines gave linear correlation coefficients $>.998$; the dotted lines represent regions
where the finite-size effect due to the boundary condition at $x=8.8$ becomes noticeable.}
\end{figure}

The system defined in (\ref{eq:grad},\ref{eq:eqnmo}) was simulated in order to test whether $\langle\delta\phi\rangle$
and $\Delta$ would exhibit the predicted power law behaviors.  As Figure \ref{fig:sim} illustrates, the concurrence of
the data with the analytical calculations is essentially exact, except near $x=8.8$, where a boundary condition
requiring $\delta\phi$ to vanish artificially skews the curve towards zero. 

We are now in the position to assess whether the gradient remains robust to changes in the average production rate
$\langle\eta\rangle=\eta_{0}$ in the presence of white noise of strength $\gamma$.  Recall that the condition for
robustness to $\eta_{0}$ in the steady-state case was $\epsilon\ll x^{*}(0)$.  Now examining the
\emph{noise-averaged} shape of the gradient, we find that
\begin{equation}
\langle\phi(x)\rangle\simeq\frac{6D/f}{(x+\epsilon)^{2}}-\frac{A}{x^{3}}
\end{equation}
where $A(\eta_{0},\gamma)$ is an unknown, positive amplitude that increases in a monotonic, unbounded fashion with
$\gamma$ from the value of zero at $\gamma=0$.
 Comparing to (\ref{eq:ss}), it is apparent the impact of the noise at leading order is that it drives up the effective
value of $\epsilon$.  This means that although our criterion for robustness may be met in the absence of noise, there
must be a $\gamma^{*}(\eta_{0})$ for which $\gamma>\gamma^{*}$ implies that
$2\epsilon(\eta_{0})+A(\gamma,\eta^{0})>x^{*}(0)$.
 Thus, the robustness of the system to changes in $\eta_{0}$ is reduced in the presence of noise, and can be eliminated
altogether if the noise level is sufficiently high.

It is also interesting to consider the effects that the fluctuations might themselves have on the process of
development. If the embryo had an infinite amount of time to sample the fluctuating level of morphogen at each point in
space, it could ``measure" and respond to the gradient $\langle\phi\rangle$ with perfect precision.  Instead, positional
information must be recovered from a morphogen gradient over the course of some finite period $T$ dictated by the
timetable of embryonic growth.  A calculation similar to the one carried out in (\ref{eq:diag}) reveals that the
variance measured at $x$ during this period will be less than $\Delta$ and asymptotically proportional to $T^{-1}x^{-6}$
(The slower $x^{-6}$ decay of fluctuations results from the fact that points more distant from the origin receive
information about the effective value of $\epsilon$ more slowly). Thus, the precision of cell type boundaries must be
limited approximately by $\Delta x^{*}\sim (x^{*})^{-3}/\langle\phi'(x^{*})\rangle$.
 This uncertainty in the boundary position should grow monotonically with $\gamma$ and depend only very weakly on
$x^{*}$.  
Thus, regardless of where the boundary lies on average, a sufficiently high noise strength could push the precision in
$x^{*}$ below tolerable levels.

The above discussion focused on a reduced model of morphogen gradient formation 
that involves only one chemical species.  In fact, even the simplest realizations of self-enhanced ligand degradation
\emph{in vivo} involve intermediary receptors and proteases that enable the morphogen to regulate its own rate of
degradation \cite{Eldar03}. However, there are at least two reasons to believe that this additional sophistication
should only magnify the effects we have observed in the reduced model.  Firstly, this study has focused on a system in
which the only source of noise was localized at the single point in space at which the damping of fluctuations was
strongest.  In contrast, more realistic models may contain multiple sources of noise that extend across the entire
system.  Secondly, in more complex pathways, it is possible for the concentration of one chemical species to control the
relaxation time of fluctuations in a second species.  Thus, a pathway that is robust to the concentration of a component
in the deterministic steady-state may exhibit fluctuations and an induced shift away from the steady-state that are not
robust to this concentration.  For example, we have recently shown that under reasonable assumptions, the gradient shape
established by the BMP pathway studied in \cite{Eldar02} should exhibit increased variability when the ambient
concentration of the short gastrulation protein is reduced.  This predicted variability is anecdotally confirmed by
experiments carried out in \cite{Sutherland03}, but the issue still demands a more quantitative investigation.

This study is, to our knowledge, the first to elucidate the impact that time-varying stochasticity in gene expression
can have on the shaping of morphogen gradients.  Using both analytical and numerical techniques, we have shown that
noise can interfere with the putatively robust specification of positional information by inducing a non-robust mean
shift in the gradient away from its steady-state shape, and can also cause fluctuations in the position of a cell fate
boundary. 
There is furthermore good reason to believe that these effects should be more pronounced in models more realistic than
the simplified system considered here.  However, while it has been made clear in this work that noise has potential
significance, it remains a question for experimenters whether fluctuations are large enough that they can play a
detectable role in the formation of morphogen gradients. Yet even in the cases where noise turns out to have little
effect, there will still remain the intriguing possibility that specific mechanisms are responsible for attenuating
fluctuations in order to prevent them from interfering substantially with the action of morphogens.  In either event,
our findings should provide motivation for an exciting new line of experimental inquiry that may contribute to a richer
understanding of the factors that play a decisive role during embryonic development.

We are grateful to Martin Howard for thoughtful comments. JLE thanks the Rhodes Trust for financial support. JC was
supported in part by the EPSRC under grant GR/R83712/01. Part of this work was carried out while JC was a member of the
Institute for Advanced Study, supported by the Ellentuck Fund.

\bibliography{myrefs}
\end{document}